\documentclass[twocolumn,aps,pra,superscriptaddress]{revtex4}
\usepackage{graphicx}
\usepackage{amsmath}
\usepackage{amssymb}
\usepackage{textcomp}
\usepackage{braket}
\usepackage[colorlinks=true,linkcolor=blue,citecolor=blue,urlcolor=blue]{hyperref}

\begin{document}

\title{Active Learning Approach to Optimization of Experimental Control}

\author{Yadong Wu}
\affiliation{Institute for Advanced Study, Tsinghua University, Beijing,
100084, China}

\author{Zengming Meng}
\affiliation{State Key Laboratory of Quantum Optics and Quantum Optics Devices, and Institute of Opto-Electronics, Collaborative Innovation Center of Extreme Optics, Shanxi University, Taiyuan 030006, China}

\author{Kai Wen}
\affiliation{State Key Laboratory of Quantum Optics and Quantum Optics Devices, and Institute of Opto-Electronics, Collaborative Innovation Center of Extreme Optics, Shanxi University, Taiyuan 030006, China}

\author{Chengdong Mi}
\affiliation{State Key Laboratory of Quantum Optics and Quantum Optics Devices, and Institute of Opto-Electronics, Collaborative Innovation Center of Extreme Optics, Shanxi University, Taiyuan 030006, China}

\author{Jing Zhang}
\email{jzhang74@sxu.edu.cn}
\affiliation{State Key Laboratory of Quantum Optics and Quantum Optics Devices, and Institute of Opto-Electronics, Collaborative Innovation Center of Extreme Optics, Shanxi University, Taiyuan 030006, China}

\author{Hui Zhai}
\email{hzhai@tsinghua.edu.cn}
\affiliation{Institute for Advanced Study, Tsinghua University, Beijing,
100084, China}

\date{\today}

\begin{abstract}
In this work we present a general machine learning based scheme to optimize experimental control. The method utilizes the neural network to learn the relation between the control parameters and the control goal, with which the optimal control parameters can be obtained. The main challenge of this approach is that the labeled data obtained from experiments are not abundant. The central idea of our scheme is to use the active learning to overcome this difficulty. As a demonstration example, we apply our method to control evaporative cooling experiments in cold atoms. We have first tested our method with simulated data and then applied our method to real experiments. We demonstrate that our method can successfully reach the best performance within hundreds of experimental runs. Our method does not require knowledge of the experimental system as a prior and is universal for experimental control in different systems.
\end{abstract}

\maketitle

\section{Introduction}

Machine learning can find broad applications in physics research. One of the major advantages of machine learning algorithms is their excellent performance for optimization. On the other hand, optimization is one of the most common tasks in physics experiments, as experimentalists always need to adjust control parameters to reach the best performance.  Therefore, the advantage of the machine learning algorithms fits perfectly the demand of experiments in physical sciences. In recent years, there are growing interests in applying various kinds of machine learning based algorithms for optimizing experimental controls \cite{control1, control2, control3, control4, control5, control6, control7, control8, control9, control10, control11, control12, control13, control14, control15, control17, control18, control19, control20, control21, control22, evap1, evap2, evap3, evap4}.

Here we focus on a class of experimental control problems as described below. First of all, the entire control process can be quantified. All the control parameters, say, denoted by $\boldsymbol{\alpha}=\{\alpha_1, \dots, \alpha_n\}$, can be quantitatively determined. And once these control parameters are fixed, the performance $\mathcal{F}$ is also quantitatively determined and measuring $\mathcal{F}$ is repeatable.  Secondly, the goal of the experimental control is unique, which is to optimize the performance $\mathcal{F}$. In other words, one needs to reach either the maximum or the minimum of $\mathcal{F}$. Thirdly, if the experimental process is not complicated and is well understood theoretically, the function relation between the control parameters ${\bm \alpha}$ and the control output $\mathcal{F}$ can be calculated theoretically, and the theoretical simulation can be used to guide the experimental optimization. However, for many cases as we consider here, because of various complicated facts in reality, it is hard to reliably determine the function theoretically, and in some cases, it is hard even to determine its general functional form. So, for each given ${\bm \alpha}$, we can only determine the performance by performing experimental measurements. Finally, the experimental measurements require resources. In many cases, it is possible to carry out hundreds of experimental runs but it is impossible to obtain large datasets including hundreds of thousands of data.

Since there exits a deterministic function relation between ${\bm \alpha}$ and $\mathcal{F}$, it is conceivable that we can empirically determine this function by fitting the data. Hence, the neural network (NN) shows its advantage here. Because of the great expressibility of the NN, a deep enough NN can express a general function without assuming its explicit form as a prior. Therefore, a natural idea is to fit the functional mapping between ${\bm \alpha}$ and $\mathcal{F}$ with a NN. Then, with this NN, we can find out the extreme of this function that gives rise to the optimal control parameter. However, there is a main obstacle to this approach. In order to learn a multi-dimensional function, usually one needs a lot of data. The number of data that can be obtained from experiments limits the use of the NN based approach. In other words, it is a major challenge of applying NN based algorithms to optimize experimental control that the total numbers of labeled data are not abundant.

However, on the second thought, one realizes that we actually do not need to fit the entire function in order to determine its extreme. In fact, we only need to know the function in the neighborhood of its extreme. Therefore, we only need to measure data in this neighborhood regime and the NN also only needs to fit the data accurately in this regime, with which we are able to determine the optimal parameters. This can certainly significantly reduce the demand of data. However, it comes to a ``paradox". On one hand, to reach our goal with the minimal number of data, we should try to sample the data points nearby the extreme. On the other hand, since our goal is to find out the extreme, we do not know where the extreme regime locates before we reach our goal.

In this work, we develop an \textit{active learning} approach to solve this paradox. Active learning is a stratagem that guides sampling data iteratively \cite{active1, active2}. In recent years, active learning has found more and more applications in physical science \cite{control1, control2, control3, control4, control5, control6, control7, control8, ALphy1, ALphy2, ALphy3, ALphy4, ALphy5, ALphy6, ALphy7, material1, material2, material3, material4, material5, material6, material7, material8, material9, evap1, evap2, evap3, evap4}, especially in computation problem for material science and chemistry \cite{material1, material2, material3, material4, material5, material6, material7, material8, material9}. Below, we will first describe our general active learning scheme for controlling experiments, and then as an example, we apply our method to optimize evaporative cooling experiments in cold atomic gases.

\begin{figure}[t]
\begin{center}
\includegraphics[width=0.5\textwidth]{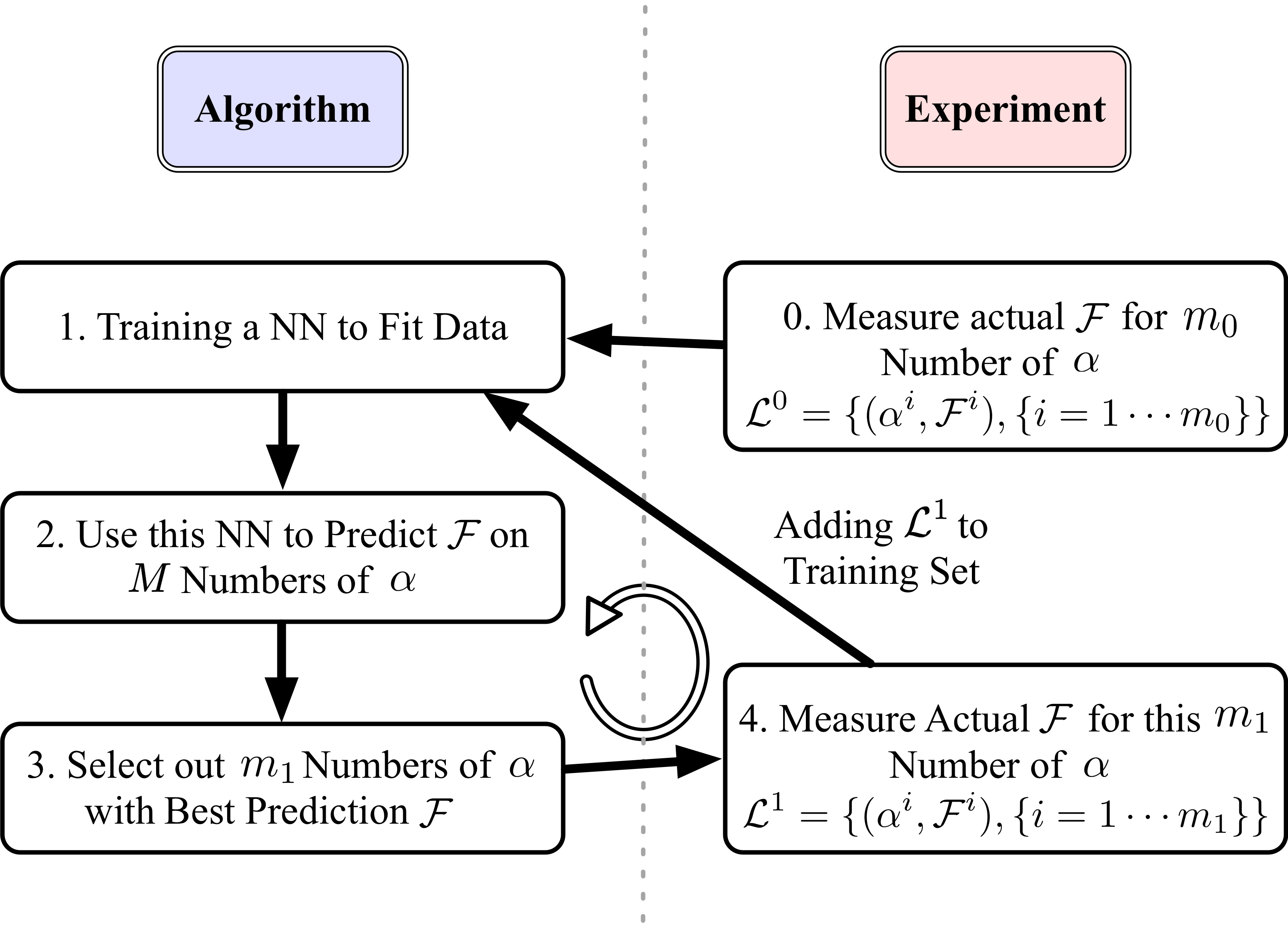}
\end{center}
\caption{Active learning protocol for optimizing experimental control. Here ``NN" represents ``neural network".  $M$ is a large number such as a few hundred thousands or a few millions, and both $m_0$ and $m_1$ are small numbers such as a few tenth. The circle with arrow indicates iteration of steps 1-4 until it converges.   }
\label{protocal}
\end{figure}

\section{General Scheme}

We present our general scheme in Fig. \ref{protocal}. The basic idea is to first train a NN with a small number of randomly sampled labelled data and then use this NN to guess where the possible extreme regime locates. This estimation guides further query of actual data, with which one refines the fitting nearby the extreme. This process continues iteratively until the prediction of the NN and the actual measurements converge in the neighborhood of the extreme. More concretely, the entire process contains the following five steps:

0. Initially, we randomly generate $m_0$ number of different control parameters ${\bm \alpha}$, and experimentally measure the performance $\mathcal{F}$ for each control. Here $m_0$ is a small number usually taken as a few tenth. This generates an initial dataset $\mathcal{L}^0=\{({\bm \alpha}^j, \mathcal{F}^j),j=1,\dots,m_0\}$. Especially, we shall pay attention to a sub-set $\bar{\mathcal{L}}^0\subset \mathcal{L}^0$ which contains $\bar{m}_0$ ($\bar{m}_0<m_0$) number of cases with the best performance.

1. We design a NN and train it with the initial dataset $\mathcal{L}^0$. Since the number of data is small, the training results can depend on the initialization. Therefore, we run different initializations and choose one with the smallest loss. Especially, we should pay special attention to the loss on the sub-set $\bar{\mathcal{L}}^0$ when we choose a NN structure and an initialization.

2. We use the trained NN to make predictions on a very large dataset with $M$ different control parameters ${\bm \alpha}$. Here $M$ is a large number, say, a few hundred thousands or a few millions. When sampling these parameters, we simultaneously take two different kinds of probability distributions. Part of the parameters are sampled with a probability obeying the Gaussian distribution around the parameters in $\bar{\mathcal{L}}^0$, and part of the parameters are sampled with a probability obeying the uniform random distribution in the entire parameter space. The latter is important to ensure that we will not miss other possible extremes that are not covered in the initial dataset. The ratio between two different ways of sampling can be adjusted as the process continues. The weight on the former can increase in the later stage when the results gradually converge.

3. We select out $m_1$ number of control parameters among all $M$ control parameters explored by the NN in the step-2, and these $m_1$ number of control parameters have the best performance $\mathcal{F}$ predicted by the NN. Here, again $m_1$ is a small number of a few tenths. However, we should keep in mind that since the NN is trained with a small dataset $\mathcal{L}^0$, the predication of this NN may not be accurate.

4. We experimentally measure the actual performance of these $m_1$ number of control parameters proposed by the NN in the step-3, and the measurements return the actual performance $\mathcal{F}$. To quantify whether the NN converges in the neighborhood of the extreme, we compare the actual value with the prediction of NN for a few best ones. If not converged, we add this $m_1$ number of dataset $\mathcal{L}^1=\{({\bm \alpha}^j, \mathcal{F}^j),j=1,\dots,m_1\}$ into the training set $\mathcal{L}^0$ and train the NN again.

\begin{figure}[t]
\begin{center}
\includegraphics[width=0.45\textwidth]{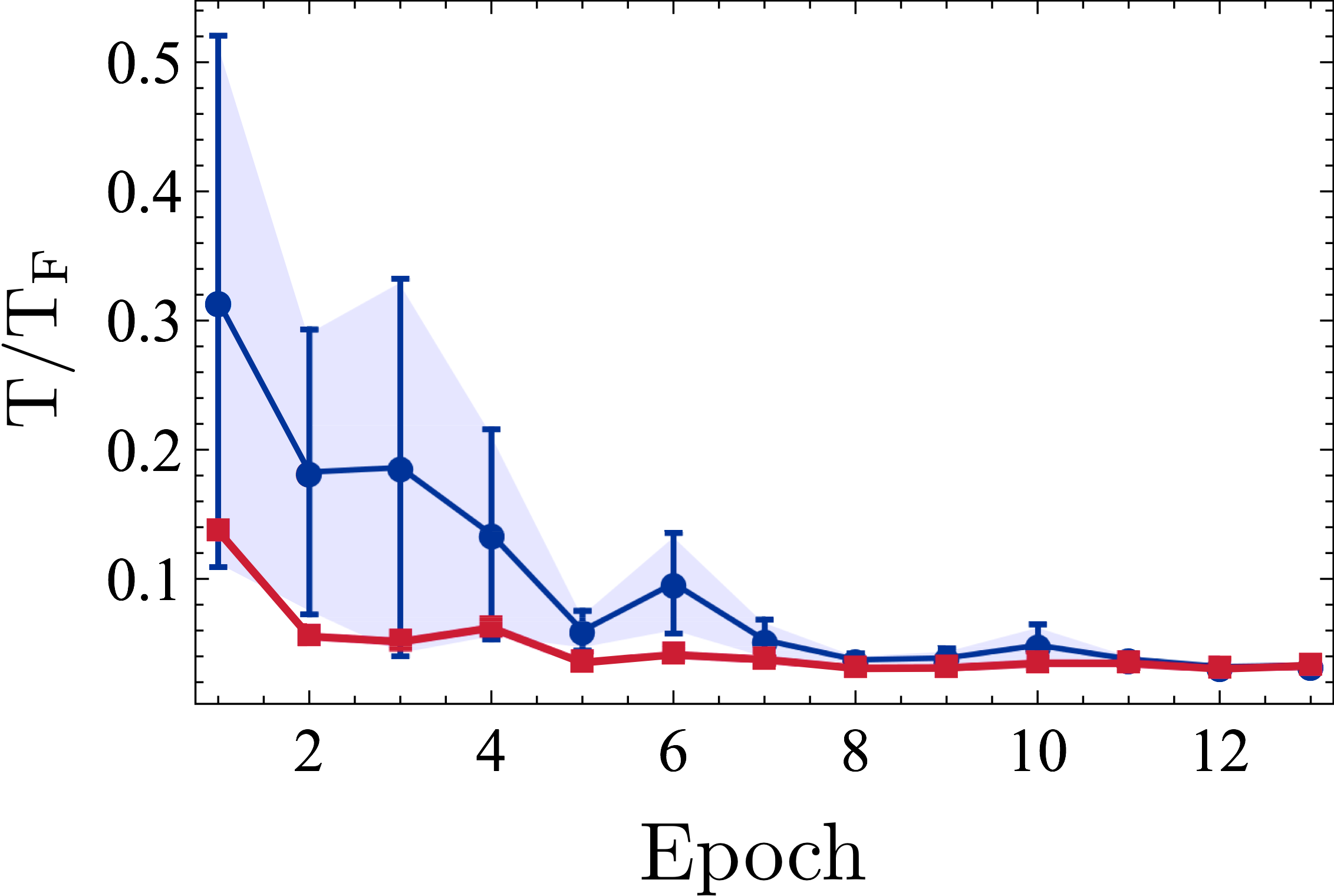}
\end{center}
\caption{The lowest $T/T_\text{F}$ as a function of training epoch $\mathcal{S}$. The horizontal axes is the number of epochs. The red point is the lowest $T/T_\text{F}$ in all $m_1$ control trajectories proposed at each epoch, and the blue point is the average of the lowest five $T/T_\text{F}$. The error bar indicates the variance of this average. Here we use the simulated results with theory provided by Ref. \cite{Jin} to determine temperature. }
\label{theory_training}
\end{figure}

It forms a circle from the step-1 to the step-4, and each of such circle is called an Epoch. We continue the circle until the prediction of NN and the actual measurements agree with each other. Note that here we do not require the agreement in the entire parameter regime, and only require the agreement in the regime around the extreme. Suppose we can reach convergence in few tenths of epochs and the number of epochs is denoted by $\mathcal{S}$. The total number of data required is $m_0+m_1\times \mathcal{S}$, which is of the order of a few hundred. In this way, we can reach the optimal experimental control with a few hundreds of runs. Here we remark that in this algorithm, the NN needs to sample a huge number of possible control parameters at the step-3 of each epoch, which causes computational resources of the NN. That is to say, this algorithm makes a trade-off between experimental resources and computational resources.

Here we should also remark that this algorithm is reminiscent of how human reaches an optimal control. Suppose a new student Bob has no prior knowledge of the experiment that he will work on and unfortunately he also has no help from others. What Bob can do is to first randomly try some controls. With this random tries, he builds up some intuitions in his mind. This intuition is an analogy of the first NN trained by the initial dataset. This intuition tells him which regime is probably more favorable for reaching optimal control and which regime is maybe unfavorable. Next, he will try more experiments in the favorable regime and fewer experiments in the unfavorable regime. These further tries correct his intuition and make it more and more accurate. When Bob has a good intuition about his experiment, it means that the prediction of his brain agrees very well with the actual measurement, at least in the regime with good performance.

\section{Example: Evaporative Cooling}

\begin{figure}[t]
\begin{center}
\includegraphics[width=0.45\textwidth]{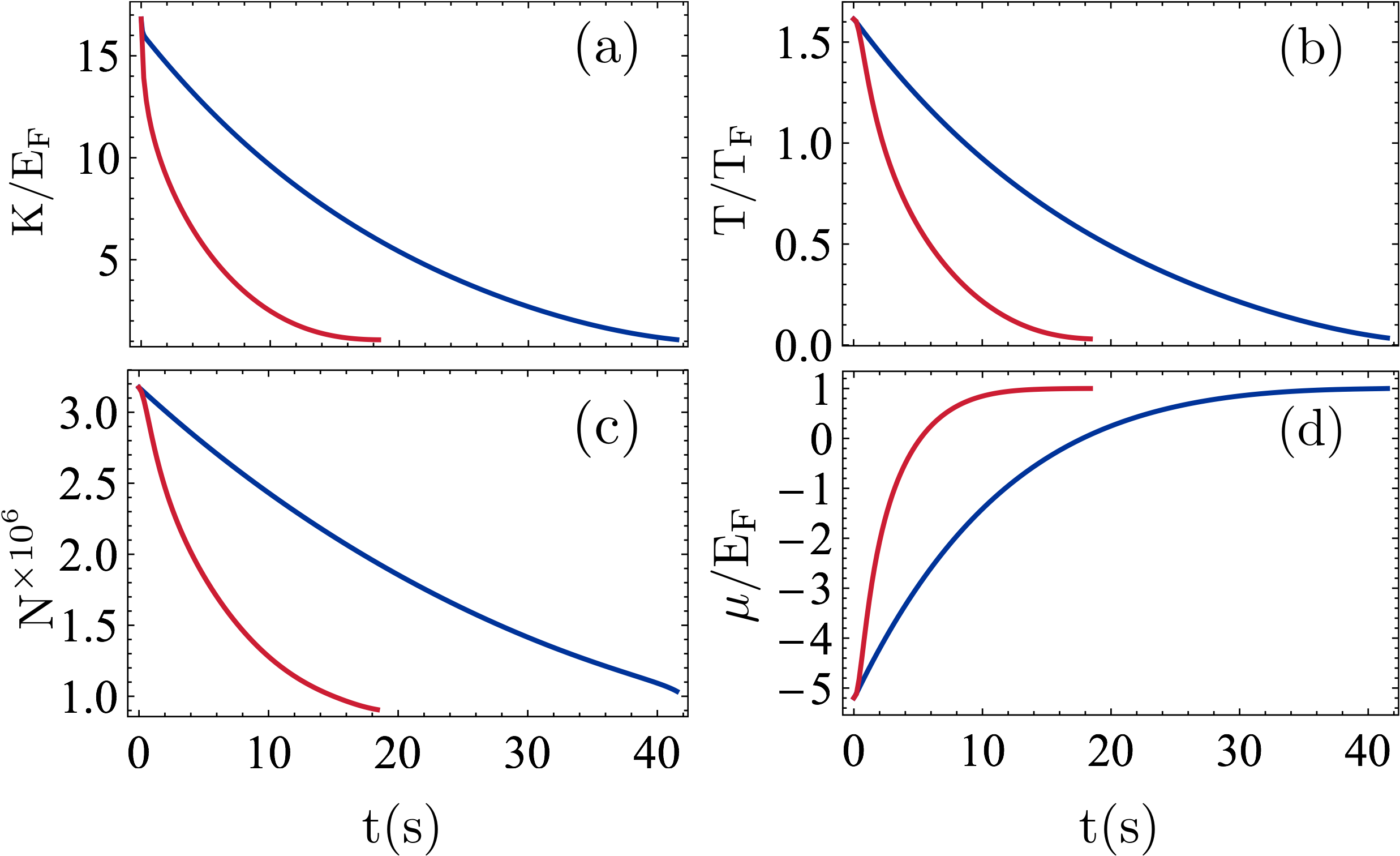}
\end{center}
\caption{The trap depth $K/E_\text{F}$ (a), the temperature $T/T_\text{F}$ (b), the total atom number $N$ (c),  and the chemical potential $\mu/E_\text{F}$ (d) as a function of evaporative cooling time. The blue lines are the optimal control found in Ref. \cite{Jin} and the red lines are the optimal control found by the active learning scheme.    }
\label{theory_curve}
\end{figure}

As a demonstration example, we apply our method to optimize evaporative cooling for cold atoms. Evaporative cooling is a universal scheme to reach quantum degeneracy for cold Bose and Fermi atomic gases \cite{early1, early2, early3, early4, early5, early6, early7, early8, early9, early10, early11, Jin2}. Cold atoms are confined in a trap created by either magnetic fields or laser fields. The basic idea is to reduce the trap depth as a function of time, such that atoms with the kinetic energy higher than the trap depth will escape from the trap. Therefore, the averaged kinetic energy of the remaining atoms will be reduced such that the temperature will be lowered after the system rethermalizes. Nevertheless, we should notice that the evaporative cooling pays the price of loss atoms in order for decreasing temperature. When the number of atoms is reduced, the characteristic temperature scale for quantum degeneracy also decreases. This characteristic temperature scale is the Bose-Einstein condensation temperature $T_\text{c}$ for bosons and the Fermi temperature $T_\text{F}$ for fermions. To characterize quantum degeneracy, what really matters is how small the ratio $T/T_\text{c}$ or $T/T_\text{F}$ one can achieve with the evaporative cooling.

\begin{figure}[t]
\begin{center}
\includegraphics[width=0.45\textwidth]{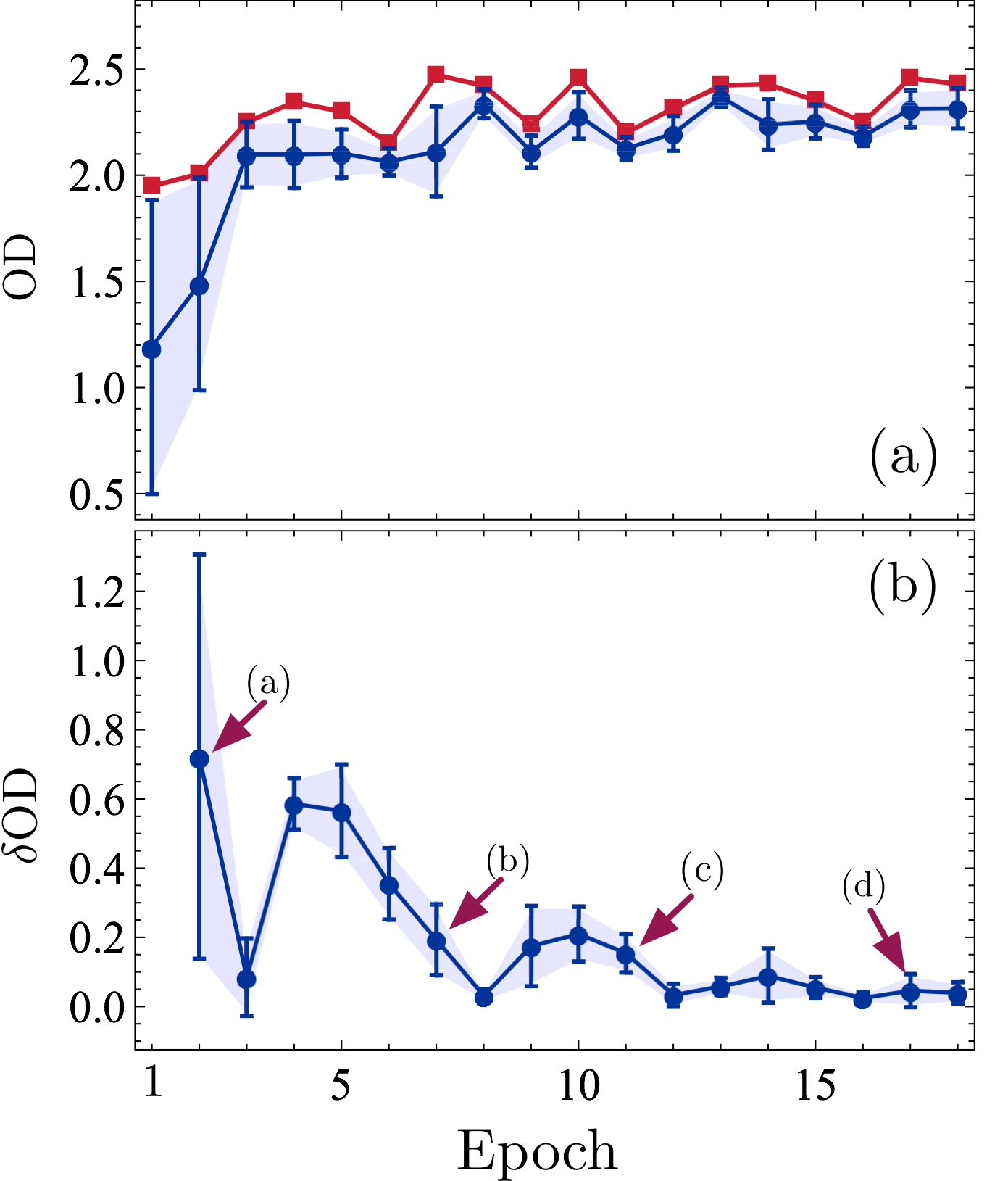}
\end{center}
\caption{(a) The dimensionless optical density (OD) of a Bose condensate. (b): The relative difference $\delta \text{OD}$ between the prediction of the NN and the actual experimental measurements. Both are plotted as a function of the training epoch. The red points in (a) are the best OD for all $m_1$ control trajectories proposed by the NN at each epoch, and the blue points are averaged results of the five highest ones. The points in (b) are also averaged over the five trajectories with the best performance. Error bars indicate the variance for taking the average. Here the actual results are obtained by experimental measurements. The OD is measured after $53$ms time-of-flight. }
\label{exp_training}
\end{figure}

Suppose we start with a fixed initial trap depth $K_0$ and an initial atoms number, what one can control during the evaporative cooling process is how the trap depth $K$ changes as a function of time $t$. We parametrize the $K(t)$ trajectory as follows
\begin{equation}
K(t)=K_0\left\{1+\sum\limits_{i=1}^{l}\left[a_i\left(\frac{t}{t_\text{f}}\right)^i+b_i\left(\frac{t}{t_\text{f}}\right)^{1/(i+1)}\right]\right\}. \label{evaporation_curve}
\end{equation}
with $t\subset[0,t_\text{f}]$. Here the control parameter ${\bm \alpha}=\{a_{i}, b_{i}, (i=1,\dots,l),t_\text{f}\}$. The evaporative cooling is therefore a typical optimization problem as we described above. We need to find the best control parameter ${\bm \alpha}$, and the control goal is to reach the lowest $T/T_\text{c}$ or $T/T_\text{F}$. Below we will apply the active learning scheme described above to optimize the evaporative cooling process. We note there are already several works using machine learning based methods to optimize evaporative cooling \cite{evap1, evap2, evap3, evap4}, but their methods are different from ours.

\subsection{Simulated Data}

Since the evaporative cooling process is a complicated non-equilibrium dynamics of an interacting many-body system, it is hard to accurately simulate this process theoretically. However, there are efforts to understand the evaporative cooling process, for instance, based on the quantum kinetic equations \cite{Jin, Jin2, LuoLe}. Here, to test our method, we first use the theoretical simulation of the evaporative cooling of two-component fermions introduced in Ref. \cite{Jin} and compute the final $T/T_\text{F}$ for each given evaporation curve Eq. \ref{evaporation_curve}. Hence, we use $T/T_\text{F}$ as $\mathcal{F}$, and the goal is to find the smallest $T/T_\text{F}$. We follow the general scheme discussed above, except for the experimental measurements are replaced by numerical simulation. Here we choose $m_0=m_1=10$.

\begin{figure}[t]
\begin{center}
\includegraphics[width=0.45\textwidth]{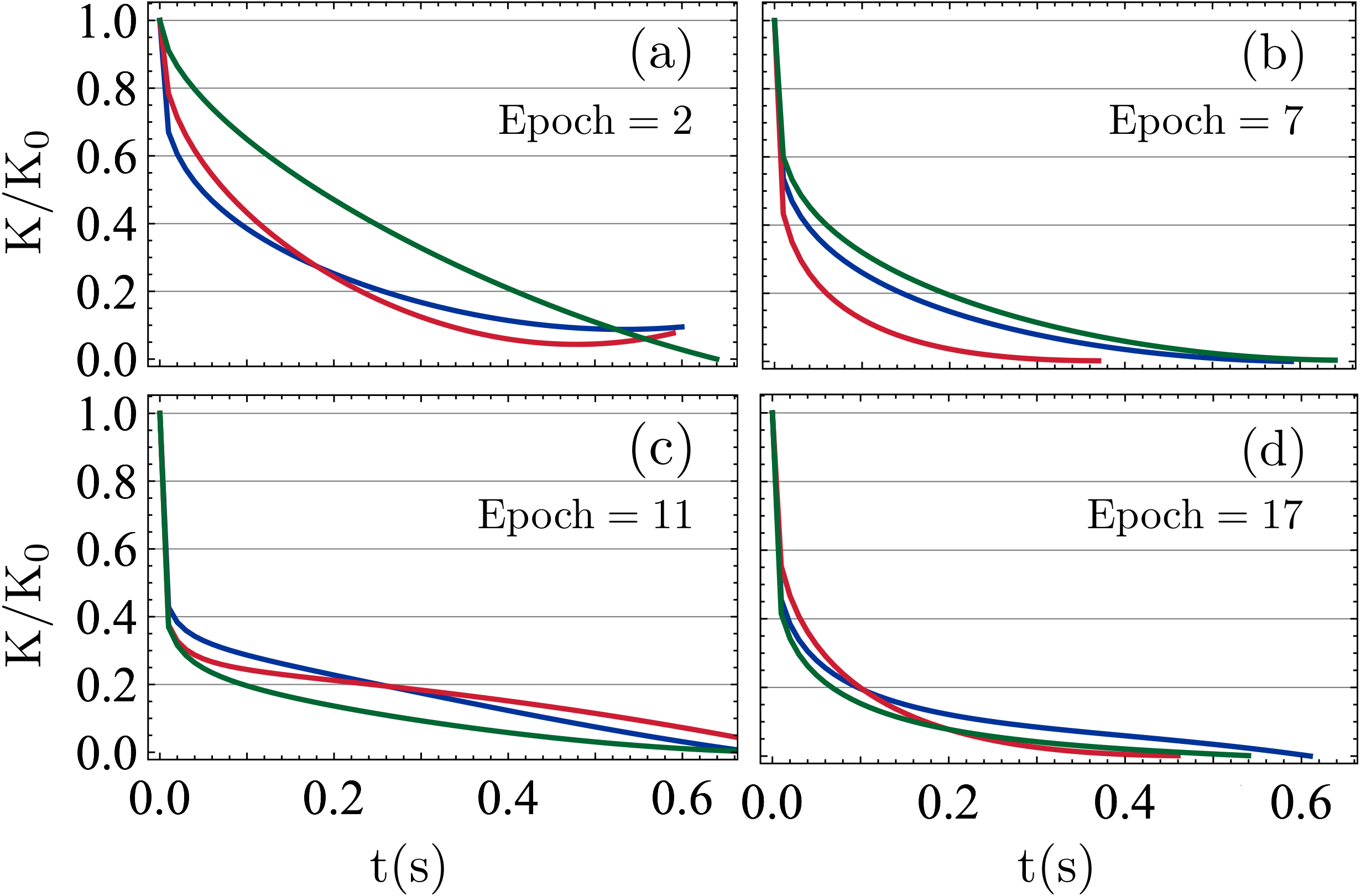}
\end{center}
\caption{The three trajectories with the highest optical density at different epochs, as marked by arrows in Fig. \ref{exp_training}(b)  }
\label{exp_curve}
\end{figure}

As we show in Fig. \ref{theory_training}, we plot $T/T_\text{F}$ as the lowest temperature, or averaged over the lowest five temperatures, among the actual values of all $m_1$ control trajectories proposed by the NN at each epoch. One can see that the temperature $T/T_\text{F}$ decreases and the variance among the five lowest temperatures is also suppressed as the training epoch increases. Eventually it achieves the lowest temperature $T/T_\text{F}\sim 0.03$. In Fig. \ref{theory_curve}, we compare our best trajectory with the one found in Ref. \cite{Jin}. Ref. \cite{Jin} proposed a classical method to optimize the trajectory that maximizes the kinetic energy removed per atom from the system at each small time interval. Fig. \ref{theory_curve} shows that although at the end of the evaporation, both two methods yield similar $T/T_\text{F}$ and atom number, these two trajectories are quite different, and the evaporation time for the trajectory obtained by our method is only about half of the trajectory obtained in Ref. \cite{Jin}.

\subsection{Experimental Data}

With the success in the test with simulated data, we now apply our method to real experiment. In this case, we consider evaporative cooling of ${}^{87}$Rb Bose gas, and the experimental apparatus is described in Ref. \cite{Zhang,Xiong2010,Chai2012}. Though one can measure temperature in cold atomic gases, it is more direct to measure the phase space density. More precisely, since the density of cold atomic gases is measured by absorption imaging of laser after the time-of-flight, a dimensionless quantity called the optical density (OD) is used for quantifying the phase space density \cite{OD1, OD2}.  Larger OD means higher phase space density, and the higher the OD, the lower $T/T_\text{c}$. Hence, we take the OD as $\mathcal{F}$ and the control goal is to maximize the OD after a sufficiently long time-of-flight. Here we take $m_0=m_1=20$.

The results are shown in Fig. \ref{exp_training}. Fig. \ref{exp_training}(a) shows the highest OD, or the average over the highest five OD, among all $20$ trajectories proposed by the NN at each epoch. Similar as Fig. \ref{theory_training}, one can see the OD increases and the variance is suppressed as the training epoch increases. In Fig. \ref{exp_training}(b) we compare the prediction by the NN $\text{OD}^\text{NN}$ with the actual measured results $\text{OD}^\text{Exp}$ for the best five trajectories in each epoch. Here $\delta \text{OD}$ is defined as $\sum\limits_{i=1}^{5}|\text{OD}_i^\text{NN}-\text{OD}_i^\text{Exp}|/\text{OD}_i^\text{Exp}$. Fig. \ref{exp_training}(b) shows $\delta \text{OD}$ approaches zero as training epoch continues. In Fig. \ref{exp_curve}, we show the three trajectories with the highest measured OD at different epoch. One can see that initially these trajectories are quite different, and they gradually converge in the later stage. These two figures show that the NN successful finds the extreme around $\sim 15$ epochs and the NN converges in the neighborhood of the extreme. The total cost of experimental measures is about $360$.

\section{Summary}

In summary we present a general NN based scheme to optimize the experimental control, and the emphasis of this scheme is that we use the active learning to reduce the query of experimental measurements to the minimum. Our scheme is quite universal that can be applied to experiments in different areas, for instance, growing samples in material science and preparing quantum states or quantum gates in quantum science. More importantly, our method does not require any knowledge of the physical system as a prior, and therefore, it can be used for controlling new experiments in the future.

\textit{Acknowledgement}. This work is supported by Beijing Outstanding Young Scientist Program (HZ), MOST(Grant No. 2016YFA0301600, 2016YFA0301602, 2018YFA0307600) and NSFC (Grant No. 11734010, 11804203).

\end{document}